\documentclass[1p]{elsarticle}
\usepackage{subcaption}
\usepackage{amsmath}
\usepackage{color}
\usepackage{cancel}
\usepackage{ulem}

\journal{Journal of \LaTeX\ Templates}

\begin{document}
{\flushright{YITP-19-124}}

\begin{frontmatter}

\title{$\Lambda\Lambda$ and N$\Xi$ interactions from Lattice QCD \\ near the physical point
}



\author[ad_YITP,ad_RIKEN]{Kenji~Sasaki}
\ead{kenjis@yukawa.kyoto-u.ac.jp}

\author[ad_YITP,ad_RIKEN,ad_UTSUKUBA]{Sinya~Aoki}
\author[ad_RIKEN,ad_iTHEMS]{Takumi~Doi}
\author[ad_RIKEN]{Shinya~Gongyo}
\author[ad_iTHEMS,ad_RIKEN]{Tetsuo~Hatsuda}
\author[ad_RCNP,ad_RIKEN]{Yoichi~Ikeda}
\author[ad_NIHONU,ad_RIKEN]{Takashi~Inoue}
\author[ad_RIKEN]{Takumi~Iritani}
\author[ad_RCNP,ad_RIKEN]{Noriyoshi~Ishii}
\author[ad_RCNP,ad_RIKEN]{Keiko~Murano}
\author[ad_RIKEN]{Takaya~Miyamoto}
\author{(HAL QCD Collaboration)}

\address[ad_YITP]{Center for Gravitational Physics, Yukawa Institute for Theoretical Physics, Kyoto University, Kyoto 606-8502, Japan}
\address[ad_RIKEN]{Quantum Hadron Physics Laboratory, RIKEN Nishina Center, Saitama 351-0198, Japan}
\address[ad_UTSUKUBA]{Center for Computational Sciences, University of Tsukuba, Tsukuba 305-8577, Japan}
\address[ad_iTHEMS]{RIKEN Interdisciplinary Theoretical and Mathematical Sciences Program (iTHEMS), Saitama 351-0198, Japan}
\address[ad_RCNP]{Research Center for Nuclear Physics (RCNP), Osaka University, Osaka 567-0047, Japan}
\address[ad_NIHONU]{Nihon University, College of Bioresource Sciences, Kanagawa 252-0880, Japan}

\begin{abstract}
The $S$-wave  $\Lambda\Lambda$ and $N \Xi$ interactions are studied on the basis of the 
 (2+1)-flavor lattice QCD simulations  close to  the physical point  
  ($m_\pi \simeq 146{\rm{MeV}}$ and $m_K \simeq 525{\rm{MeV}}$).
  Lattice QCD potentials in four different spin-isospin channels are extracted 
 by using  the coupled-channel HAL QCD method and are parametrized by
  analytic functions   to calculate the scattering phase shifts.
   The  $\Lambda \Lambda$ interaction at low energies shows only a weak attraction, which 
     does not provide a bound or resonant dihyperon.
    The  $N\Xi $ interaction in  the spin-singlet and isospin-singlet channel 
    is most attractive and lead  the $N\Xi$ system near unitarity. 
      Relevance    to the strangeness=$-2$ hypernuclei as well as  to
      two-baryon correlations in proton-proton, proton-nucleus and nucleus-nucleus collisions  is also discussed.
\end{abstract}


\end{frontmatter}


\newpage

\section{Introduction}

The baryon-baryon interactions in the strangeness ${\cal{S}}=-2$ sector attract much attention to understand the nature of the 
dihyperon \cite{Jaffe:1976yi,Sakai:1999qm,Kim:2013vym},  the structures of double-$\Lambda$ or $\Xi$ 
 hypernuclei \cite{Nakazawa:2015joa,Millener1985,Hiyama:2018lgs}, and  the two-particle correlations in pp, pA and AA collisions
 \cite{STAR,Acharya:2018gyz,Acharya:2019sms}. 
Although various models with phenomenological parameters have been proposed so far for the hyperon interactions, it is of crucial importance at present to derive  the ${\cal{S}}=-2$ interaction from first principle lattice QCD simulations. Such an attempt became possible by the development of high performance computing facilities as well  as the  theoretical progress of the HAL QCD method \cite{Ishii:2006ec,Aoki:2009ji,HALQCD:2012aa}.
  
  In this paper, we report  the (2+1)-flavor lattice QCD results of the low-energy scattering of 
   $N\Xi$ and $\Lambda\Lambda$ systems at nearly physical point ($m_{\pi} \simeq 146$ MeV and $m_{K} \simeq 525$ MeV) with a large  spacetime volume  $(L \simeq 8.1 {\rm{fm}})$ on the basis of the coupled-channel HAL QCD method \cite{Aoki:2011gt,Aoki:2012bb,Sasaki:2015ifa}.   We note that the results for the
   $N \Omega$ (${\cal{S}}=-3$) system  \cite{Iritani:2018sra} and  $\Omega \Omega$ (${\cal{S}}=-6$) system \cite{Gongyo:2017fjb}   have been recently reported with the same lattice setup.

  The organization of this paper is as follows.
   In Sec.~\ref{sec:cc}, we briefly review the coupled-channel HAL QCD method.
   In Sec.~\ref{sec:setup}, our setup  of lattice QCD simulations is summarized.   
   In Sec.~\ref{sec:num}, numerical results of the $\Lambda\Lambda$ and $N\Xi$ potentials are  presented.
  After fitting the numerical data of the lattice QCD potentials by analytic functions in Sec.~\ref{sec:fit},  
  we discuss the scattering observables such as the scattering phase shifts and inelasticity in Sec.~\ref{sec:scatt}.
    Sec.~\ref{sec:sum} is devoted to summary and concluding remarks.

\section{Coupled-channel baryon-baryon interaction}
\label{sec:cc}

In the coupled-channel HAL QCD method \cite{Aoki:2011gt,Aoki:2012bb,Sasaki:2015ifa}, baryon-baryon interactions are expressed by an energy independent potential $U^{c}_{\ c'} (\vec{r}, \vec{r}')$,  which reproduces the scattering phase shifts subject to the QCD Lagrangian.
 Here the channel $c$  ($c'$) denotes the system of two particles $c_1$ and $c_2$ ($c'_1$ and $c'_2$) with 
  the rest masses $m_{c_1}$ and $m_{c_2}$ ($m_{c'_1}$ and $m_{c'_2}$), respectively.
Let us start with a function $R^c_{\ d}$ which is defined as a normalized four-point correlation between channels $c$ and $d$;  
\begin{eqnarray}
R^{c}_{\ d}(\vec{r},t) &\equiv& \frac{\sum_{\vec{x} \in V} \langle 0 \mid B_{c_1}(\vec{r}+\vec{x}, t) B_{c_2}(\vec{x}, t) \overline{{\cal J}}_d(0) \mid 0 \rangle}{\sqrt{Z_{c_1}} \sqrt{Z_{c_2}} \exp[-(m_{c_1}+m_{c_2}) t]} 
\nonumber \\
&=& 
 \sum_{j}   \psi^{c}_{W_j} (\vec{r}) e^{-\Delta W_j^c t} A_d^{W_j} +\cdots,
\label{EQ.Rdefine}
\end{eqnarray}
where $B_{c_1}$ and $B_{c_2}$ are  local interpolating operators for the baryons, while
 ${{\cal J}}_d(0)$ is a source operator of two baryons at $t=0$.  
 The wave function renormalization factors for single baryons are denoted by
   $Z_{c_1}$ and $Z_{c_2}$. 
 The $j$-th eigen energy of the total system is denoted by $W_j$, while
 the energy shift from the threshold of the channel $c$ 
 is defined by $ \Delta W_j^c =W_j - (m_{c_1}+m_{c_2})$.
   The overlap factor of the source operator to the $j$-th eigen state is given by
   $A_d^{W_j} = \langle W_j \vert  \overline{{\cal J}}_d(0) \vert 0 \rangle$.
    The Nambu-Bethe-Salpeter (NBS) wave function with total energy $W_j$ in the channel $c$ is
    denoted by $\psi^{c}_{W_j} (\vec{r})$.    The ellipses in Eq.~(\ref{EQ.Rdefine}) 
    corresponds to the inelastic contributions beyond the elastic scatterings in the coupled channel space.

An energy-independent non-local potential can be defined through the partial differential equation
 satisfied by  $R^{c}_{d}(\vec{r},t)$ (see \cite{Sasaki:2015ifa} for details):
\begin{eqnarray}
&&
\left( \frac{1+3\delta_c^{2}}{8\mu_c} \frac{\partial^2}{\partial t^2} -\frac{\partial}{\partial t} + \frac{\nabla^2}{2 \mu_c} \right) {R^{c}_{\ d}}(\vec r, t)  
= \sum_{c'} \int d^3 r^\prime {U}^{c}_{\ c'}(\vec r, \vec r^{\,\prime}) {\Delta}^{c}_{\ c'}
{R^{c'}_{\ d}}(\vec r^{\,\prime}, t),
\label{EQ.TDHAL_CC}
\end{eqnarray}
where $\mu_c= m_{c_1} m_{c_2}/(m_{c_1}+m_{c_2})$  and $\delta_c \equiv (m_{c_1}-m_{c_2}) / (m_{c_1}+m_{c_2})$. 
 The factor ${\Delta}^{c}_{\ c'} \equiv \exp[-(m_{c'_1}+m_{c'_2}- m_{c_1}-m_{c_2})t]$ 
   plays a role to compensate the threshold energy difference  between  channels $c$ and  $c'$.
   We note that the term with second-order time-derivatives in Eq.~(\ref{EQ.TDHAL_CC}) represents
     the relativistic effect. The terms with higher-order time-derivatives are neglected,
     since those contributions are numerically negligible.
     The NBS wave functions  $\psi^{c}_{W_j} (\vec{r})$ are in general 
 not orthogonal to each other, so that   the potential matrix 
$ {U}^{c}_{\ c'}(\vec r, \vec r^\prime)$ is not necessarily Hermitian.
  Nevertheless, the energy eigenvalues $W_j$ are real by construction.

In the ${\cal{S}}=-2$ channel for the octet baryons, there are four asymptotic states from below, $\Lambda\Lambda$, $N\Xi$, 
$\Lambda \Sigma$ and $\Sigma \Sigma$.  We consider only the two low-lying scattering states
throughout this paper, so that
  we have two-by-two potential $U^{c}_{\ c'}(\vec r,\vec r')$ with  $c$ and $c'$ being either $\Lambda\Lambda$ or $N\Xi$.  
  All the   inelastic effects outside of this two-by-two coupled channel space are included implicitly in $U^{c}_{\ c'}$ 
  as long as $W_j$ stays below the $\Lambda \Sigma$ threshold~\cite{Aoki:2011gt,Aoki:2012bb}. 
 
  We  use the following local interpolating operator for octet baryons,
\begin{eqnarray}
B(x)= \epsilon^{\alpha \beta \gamma} (q_{1,\alpha}^T(x)  C\gamma_5 q_{2,\beta}(x)) q_{3,\gamma} (x) \equiv [q_1q_2]q_3,
\label{eq:singleB}
\end{eqnarray}    
where $\alpha$, $\beta$ and $\gamma$ are color indices, and   $q_1$, $q_2$ and $q_3$ takes either $u$, $d$ or $s$. 
Then the interpolating operators relevant to our analysis are  
 $B_{\Lambda}(x)=([sd]u+[us]d-2[du]s)/\sqrt{6}$,
 $B_{p}(x)= [ud]u$,    $B_{n}(x)= [ud]d$,   $B_{\Xi^0}(x)= [su]s$ and  $B_{\Xi^-}(x)= [sd]s$.   
 The source operator ${\cal J}_d(0)$ is defined by the product of the baryon operator 
 Eq.~(\ref{eq:singleB}) with $q_i(x)$ replaced by $Q_i(0) =   \sum_{\vec{x}} q_i(\vec{x},0)$ i.e.
  ${\cal J}_d = \left( [Q_1Q_2]Q_3 \right)_{d_1}  \left( [Q_{1'}Q_{2'}]Q_{3'} \right)_{d_2} $ \cite{Sasaki:2015ifa}. 
 Such a wall source  is known to have a large overlap with low-lying states~\cite{Iritani:2018zbt,Iritani:2018vfn}.

In order to handle the non-locality of the potential in Eq.~(\ref{EQ.TDHAL_CC}), we employ the derivative expansion scheme
 \cite{Aoki:2009ji,Aoki:2012bb};
\begin{eqnarray}
{U^{c}_{\ c'}}(\vec r,\vec r^{\,\prime}) 
  = \left(V^c_{\ c'}(\vec r)+ \sum_{n=1}^{\infty} V^{c\, (n)}_{\ c'}(\vec r) \nabla^n + \cdots \right)\delta(\vec r-\vec r^{\,\prime}) .
\end{eqnarray}
In this work we consider only the leading order potential, $V^c_{\ c'}(\vec r)$:
The validity of such truncation at low energies  can be checked by the $t$-dependence of the resultant potentials.
 See also Ref.~\cite{Iritani:2018zbt}  for an explicit construction of the higher order terms
 as well as  the convergence test of the derivative expansion.

\section{Simulation setup of (2+1)-flavor QCD}
\label{sec:setup}

\begin{table}[t]
\caption{Baryon masses $m_{_B}$ are given in lattice unit and GeV unit. Statistical and systematic errors are shown in the first and second parentheses, respectively. The systematic errors are estimated by the difference between the
 results obtained by the fit range $[t/a]_{\rm min.}  \le t/a \le  [t/a]_{\rm max.}$ in the Table and those by the range
 $[t/a]_{\rm min.} +2 \le t/a \le  [t/a]_{\rm max.}+2$.
 The third parentheses given in GeV unit correspond to errors from the uncertainty in the lattice cutoff, $a^{-1}$.
 }
\begin{center}
\begin{tabular}{c|ll|c}
\hline \hline
 & \multicolumn{2}{|c|}{mass} & fit range \\
Baryon & \multicolumn{1}{c}{[$m_{_B}/a$]} & \multicolumn{1}{c|}{[GeV]} &  [$t/a$] \\
\hline
$N$ &       0.40949(78)(52) & 0.9553(18)(12)(74) & $13-17$ \\
$\Lambda$ & 0.48856(49)(9)  & 1.1398(11)(2)(88)  & $15-20$ \\
$\Sigma$ &  0.52365(37)(64) & 1.2217(9)(15)(94)  & $15-20$ \\
$\Xi$ &     0.58087(51)(3)  & 1.3552(12)(1)(105) & $20-25$ \\
\hline \hline
\end{tabular}
\end{center}
\label{TAB.HADRON_MASS}
\end{table}
Lattice QCD simulations with the lattice spacing $a$ are performed on gauge configurations in the large volume ($(L/a)^4 = 96^4$),  generated in the (2+1)-flavor lattice QCD with the Iwasaki gauge action at $\beta = 1.82$ and the non-perturbatively ${\cal{O}}(a)$-improved Wilson quark action, together with the stout smearing, at nearly physical quark masses ~\cite{Ishikawa:2015rho} 
corresponding to $m_{\pi} \simeq 146 $MeV and $m_K \simeq 525$ MeV.
The lattice cutoff is $a^{-1} = 2.333(18)$GeV ($a = 0.0846(7)$fm)~\cite{Ishikawa:2015rho,Ishikawa:2018rew}
corresponding to $L \simeq 8.1$fm in the physical unit.
This is sufficiently large to accommodate the interaction potential between two particles.
For quark fields, we adopt the periodic boundary condition in the spacial direction, while the Dirichlet boundary condition in the temporal direction is imposed at $t=t_0 + 48a$ with $t_0$ being the source time. The latter   allows us to average over the forward propagation and backward propagation in time due to  time-reversal and charge conjugation symmetries.
The quark propagators from the wall source with the Coulomb gauge fixing are calculated by the domain-decomposed solver 
\cite{Boku:2012zi,Terai:2013,Nakamura:2011my,Osaki:2010vj}, and then the efficient algorithm developed 
in \cite{Doi:2012xd} is employed to calculate correlation functions. 
See also Ref.~\cite{Nemura:2015yha} for discussions of a computational algorithm.

By utilizing the hypercubic symmetry on the lattice ($4$ rotations and $96$ source locations) with $414$ gauge configurations, the total number of measurements becomes $414_{\rm{[confs]}} \times 4_{\rm{[rot.]}} \times 96_{\rm{[src.]}}$.
We use the jackknife method with 23 jackknife samples and thus 
the bin size of  $(414/23)\times 4\times 96$ data,  to estimate the statistical errors.
Table~\ref{TAB.HADRON_MASS} gives octet baryon masses assuming a 1-state fit, which are a few $\%$ heavier than the physical values 
due to slightly heavier quark masses.
We have checked that a 2-state fit by including the  data for small $t/a$ provides  
baryon masses with less than  0.4\% deviation from the results of the 1-state fit. 

\section{Numerical results of $\Lambda\Lambda$ and $N\Xi$ potentials}
\label{sec:num}

To characterize the $S$-wave $\Lambda\Lambda$ and $N\Xi$ interactions,
we use  the notation ${^{2I+1, 2s+1}S_J}$ where $I$, $s$, and $J$ stand for
 the total isospin, the total spin, and the total angular momentum, respectively.
 There is a channel coupling between  $\Lambda\Lambda$ and $N\Xi$
  in  ${^{11}S_0}$,  
   while only  $N\Xi$ contributes to  the other states,
  ${^{31}S_0}$, ${^{13}S_1}$, and  ${^{33}S_1}$.
 Note also that the obtained (central) potentials implicitly contain the effect of the tensor interactions
   in the case of ${^{13}S_1}$ and  ${^{33}S_1}$ channels.

\begin{figure}[t]
  \begin{tabular}{c}
    \begin{subfigure}{0.48\textwidth}
      \includegraphics[scale=0.35, bb=0 0 390 285]{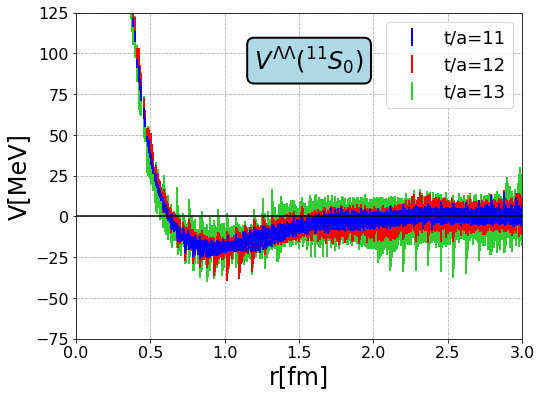}
      \caption{}
    \end{subfigure} 
    \begin{subfigure}{0.48\textwidth}
      \includegraphics[scale=0.35, bb=0 0 390 285]{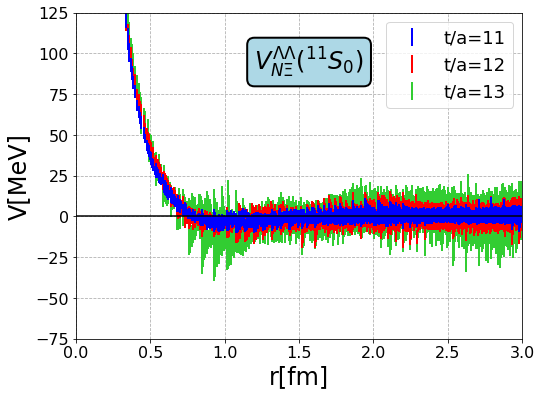} 
      \caption{}
    \end{subfigure} \\
    \begin{minipage}{1\textwidth}
    \vspace*{14mm}
    \end{minipage} \\
    \begin{subfigure}{0.48\textwidth}
      \includegraphics[scale=0.35, bb=0 0 390 285]{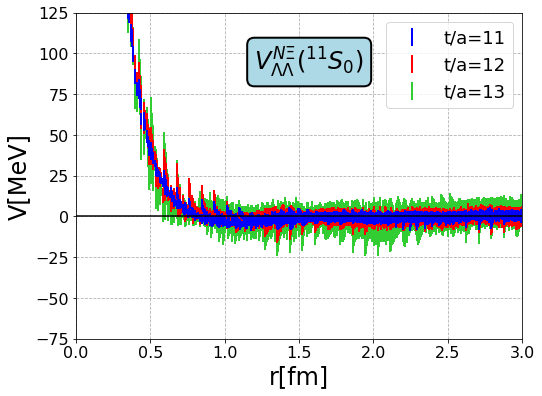}
      \caption{}
    \end{subfigure} 
    \begin{subfigure}{0.48\textwidth}
      \includegraphics[scale=0.35, bb=0 0 390 285]{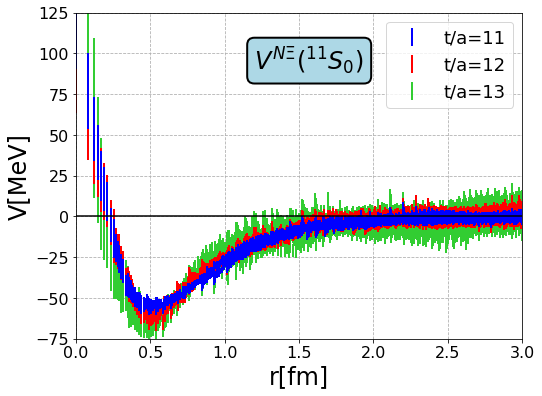}
      \caption{}
    \end{subfigure}
  \end{tabular}
\caption{The $S$-wave coupled-channel $\Lambda \Lambda$-$N\Xi$ potential in ${^{11}S_0}$.  The $V^{\Lambda \Lambda}$, 
$V^{\Lambda \Lambda}_{N \Xi}$, $V^{N \Xi}_{\ \Lambda \Lambda}$ and $V^{N \Xi}$ potentials are shown in (a), (b), (c) and (d), respectively. \label{FIG_NXS0I0}}
\end{figure}

 In Fig.~\ref{FIG_NXS0I0}, the coupled channel potentials $V^{c}_{\ c'} ({^{11}S_0})(r)$ 
 in the interval $11 \le t/a \le 13$ are shown.
   (For the diagonal part ($c=c'$),  we omit the suffix $c'$ for simplicity.)  
 See Appendix A  for the examples  of wider range of $t/a$.
     Within the statistical errors,  no significant $t$-dependence is found, which  
 implies that the leading-order  truncation of the derivative expansion is reasonable. 
 The diagonal potentials, $V^{\Lambda \Lambda}$ and $V^{N \Xi}$ in
  Fig.~\ref{FIG_NXS0I0}~(a,d), have attractive pocket with a long-range tail
 together with a short-range repulsive core.  From the meson exchange picture, the one-pion exchange is allowed only in $N \Xi$-$N \Xi$ channel.
  One interesting feature is that 
  the overall attraction in $V^{N \Xi}$ is substantially larger than that in  $V^{\Lambda \Lambda}$.
  The off-diagonal potentials shown in Fig.~\ref{FIG_NXS0I0}~(b,c) are found to be non-zero only 
 at short distance, which suggests that the $\Lambda\Lambda$-$N\Xi$ coupling is weak at low energies.

\begin{figure}[t]
  \begin{tabular}{c}
    \begin{subfigure}{0.48\textwidth}
      \includegraphics[scale=0.35, bb=0 0 390 285]{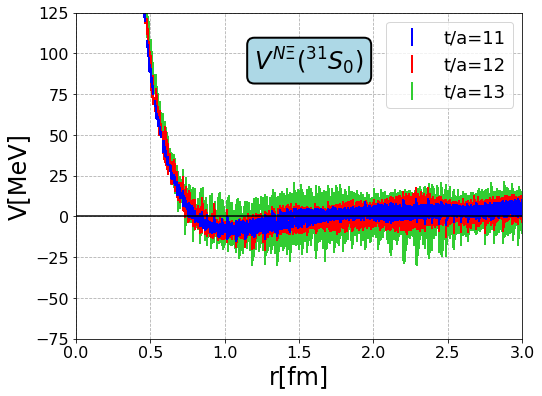}
      \caption{}
    \end{subfigure} 
    \begin{subfigure}{0.48\textwidth}
    \end{subfigure} \\
    \begin{minipage}{1\textwidth}
    \vspace*{14mm}
    \end{minipage} \\
    \begin{subfigure}{0.48\textwidth}
      \includegraphics[scale=0.35, bb=0 0 390 285]{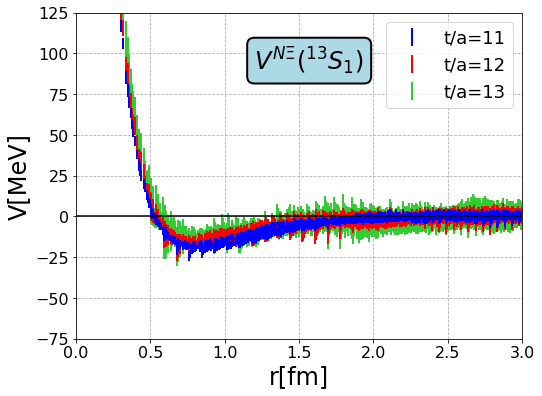}
      \caption{}
    \end{subfigure} 
    \begin{subfigure}{0.48\textwidth}
      \includegraphics[scale=0.35, bb=0 0 390 285]{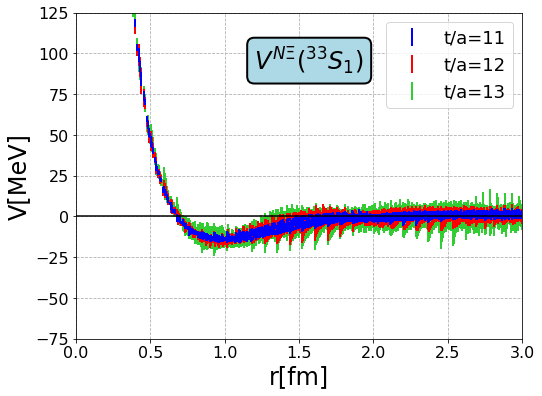}
      \caption{}
    \hspace*{1mm}
    \end{subfigure}
  \end{tabular}
\caption{The $S$-wave $N\Xi$ potentials in ${^{31}S_0}$, ${^{13}S_1}$, ${^{33}S_1}$ are plotted in (a), (b) and (c), respectively. \label{FIG_NXpot_lat}}
\end{figure}

The $S$-wave $N \Xi$ potentials in the ${^{31}S_0}$, ${^{13}S_1}$ and ${^{33}S_1}$ states are shown in Fig.~\ref{FIG_NXpot_lat} (a), (b) and (c), respectively.
Again, no significant  $t$-dependence of the potentials is found in the interval $t/a=11-13$ within the statistical errors.
  Also,  they have  stronger repulsive core and weaker mid-range attraction than those in the ${^{11}S_0}$  $N\Xi$ potential.

\begin{figure}[t]
  \begin{tabular}{c}
    \begin{subfigure}{0.48\textwidth}
      \includegraphics[scale=0.35, bb=0 0 390 285]{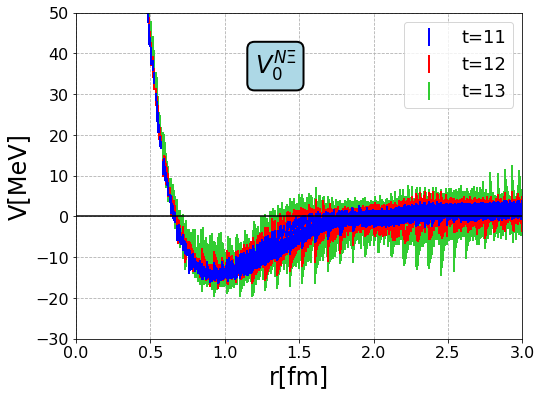}
      \caption{}
    \end{subfigure} 
    \begin{subfigure}{0.48\textwidth}
      \includegraphics[scale=0.35, bb=0 0 397 284]{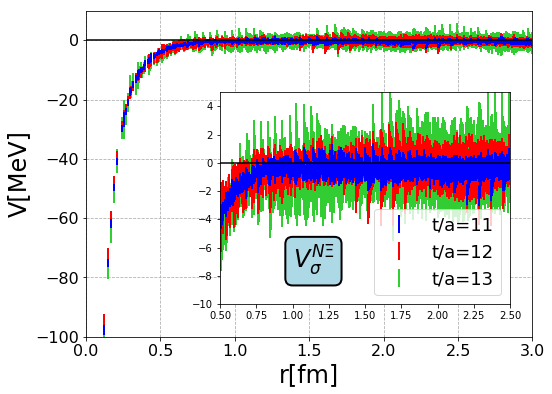}
      \caption{}
    \end{subfigure} \\
    \begin{minipage}{1\textwidth}
    \vspace*{13mm}
    \end{minipage} \\
    \begin{subfigure}{0.48\textwidth}
      \includegraphics[scale=0.35, bb=0 0 387 285]{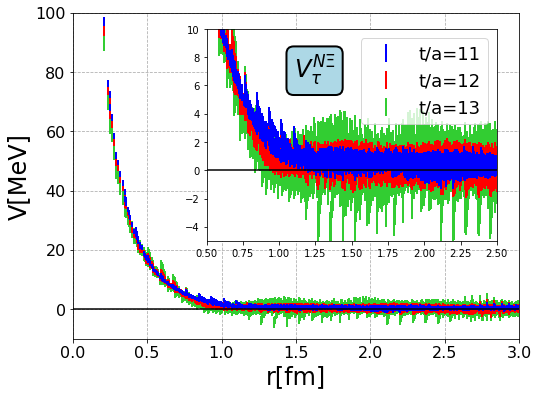}
      \caption{}
    \end{subfigure} 
    \begin{subfigure}{0.48\textwidth}
      \includegraphics[scale=0.35, bb=0 0 397 284]{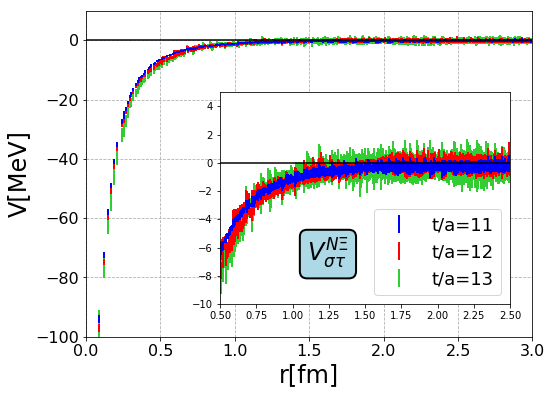}
      \caption{}
    \end{subfigure}
  \end{tabular}
\caption{The $S$-wave $N\Xi$ potentials in the operator basis: $V_0$, $V_\sigma$, $V_\tau$ and $V_{\sigma \tau}$ are shown in (a), (b), (c) and (d), respectively. The inset shows a zoom-up  in the interval, $0.5\ {\rm fm} \le r \le 2.5\ {\rm fm}$.\label{FIG_NX_op}}
\end{figure}

To capture the strong spin and isospin dependence of the 
 $N\Xi$ potentials,   following decomposition  with the operator basis is useful  \cite{Okubo:1958aa}
\begin{eqnarray}
V(r) 
= V_0(r) + V_\sigma(r) (\vec{\sigma}_1 \cdot \vec{\sigma}_2) 
+ V_\tau(r) (\vec{\tau}_1 \cdot \vec{\tau}_2) + V_{\sigma \tau}(r) (\vec{\sigma}_1 \cdot \vec{\sigma}_2) (\vec{\tau}_1 \cdot \vec{\tau}_2).
\label{eq:op_basis}
\end{eqnarray}
This is equivalently rewritten as  a relation between the spin-isospin basis and the operator basis;
\begin{eqnarray}
\left( \begin{array}{c} 
V({^{11}S_0})(r) \\ 
V({^{31}S_0})(r) \\ 
V({^{13}S_1})(r) \\ 
V({^{33}S_1})(r) 
\end{array} \right)
= \left( \begin{array}{cccc} 
  1 & -3 & -3 &  9 \\ 
  1 & -3 &  1 & -3 \\ 
  1 &  1 & -3 & -3 \\ 
  1 &  1 &  1 &  1
\end{array} \right)
\left( \begin{array}{c} 
V_{0}(r) \\ V_{\sigma}(r) \\ V_{\tau}(r) \\ V_{\sigma \tau}(r) 
\end{array} \right)
\equiv 
\hat{A} 
\left( \begin{array}{c} 
V_{0}(r) \\ V_{\sigma}(r) \\ V_{\tau}(r) \\ V_{\sigma \tau}(r) 
\end{array} \right).
\label{EQ:rotation}
\end{eqnarray}

Shown in Fig.~\ref{FIG_NX_op} are the $N \Xi$ potentials in the operator basis.
The scalar part of the $N \Xi$ potential, $V_0^{N \Xi}$, have an attractive pocket at around $1.0$ fm as
well as the short-range repulsion. The former may be related to the correlated two-pion exchange 
 as in the case of the mid-range attraction in the $S$-wave  $NN$ interactions.
We also find that $V_{\sigma \tau}^{N \Xi}$ has a long-range attractive tail, which  is consistent with the 
 one-pion exchange picture.

\section{Analytic forms of $\Lambda\Lambda$ and $N\Xi$ potentials}
\label{sec:fit}

For phenomenological applications,  it is useful to fit the LQCD potential in terms of  a combination
 of simple  analytic  functions.

 For the diagonal $\Lambda \Lambda$-$\Lambda \Lambda$ potential and the off-diagonal 
 $\Lambda \Lambda$-$N\Xi$ potential in the ${^{11}S_0}$ channel
 shown in Fig.~\ref{FIG_NXS0I0} (a,b),  we consider the following fit functions,
\begin{eqnarray}
\label{eq:YD}
V^{\Lambda\Lambda}(r) 
 &=& \sum_{i=1}^2 \alpha_i^{\Lambda\Lambda} e^{-\frac{r^2}{{\beta_i^{\Lambda\Lambda}}^2}} + \lambda_2^{\Lambda\Lambda} \left[ {\cal{Y}}(\rho_2^{\Lambda\Lambda}, m_\pi, r) \right] ^2, \\
\label{eq:YS}
V^{\Lambda\Lambda}_{N\Xi}(r) 
 &=& \sum_{i=1}^2 \alpha_i e^{-\frac{r^2}{{\beta_i}^2}} + \lambda_1 {\cal{Y}}(\rho_1, m_K, r),
\end{eqnarray}
where  the Yukawa function with a form factor ${\cal{Y}}$ is defined as 
\begin{eqnarray}
{\cal{Y}}(\rho, m, r) \equiv \left( 1 - e^{-\frac{r^2}{\rho^2}} \right) \frac{e^{-mr}}{r}.
\end{eqnarray}
In Eqs.~(\ref{eq:YD},\ref{eq:YS}), the Gauss functions describe  the short range part of the potential, 
while the Yukawa functions are motivated by the meson exchange picture 
  at medium and long range distances.
   In particular, the squared Yukawa function in eq.~(\ref{eq:YD}) 
 represents the two-pion process
 in the    $\Lambda \Lambda - \Lambda \Lambda$ interaction whose long-range part does not 
  exchange   isospin and strangeness.  Similarly, 
  the  Yukawa function in eq.~(\ref{eq:YS}) 
 represents the longest range single-kaon process
 in  the  $\Lambda \Lambda - N \Xi$ transition. 
  Note that  the kaon and pion masses
 $m_K$ and $m_{\pi}$
 are fixed to be the measured values on the lattice,  $525$  MeV and $146$ MeV, respectively.

As for the  fitting to the $N \Xi$ potentials  in Fig.~\ref{FIG_NXS0I0} (d) and Fig.~\ref{FIG_NXpot_lat} (a,b,c), 
we consider the following analytic form;
 \begin{eqnarray}
V(C)(r) 
= \sum_{i=1}^3 \alpha_i(C) e^{-\frac{r^2}{{\beta_i}^2}}
+ \lambda_2(C) \left[ {\cal{Y}}(\rho_2, m_\pi, r) \right]^2 
+ \lambda_1(C) {\cal{Y}}(\rho_1, m_\pi, r) ,
\label{V(C)}
\end{eqnarray}
with $C$ being  ${^{11}S_0}$, ${^{31}S_0}$, ${^{13}S_1}$ or ${^{33}S_1}$.
The range parameters $\beta_{1,2,3}$ and $\rho_{1,2}$ for $N\Xi$ potentials are assumed to be independent of $C$.
The above form is motivated by  the following analytic forms in the operator basis where 
   the one-pion and two-pion exchange contributions are singled out 
 explicitly in  $V_{\sigma \tau}$ and $V_0$, respectively;
 \begin{eqnarray}
V_0(r) &=& \sum_{i=1}^3 \alpha_i^{(0)} e^{-\frac{r^2}{{\beta_i}^2}}
+ \lambda_2^{(0)} \left[ {\cal{Y}}(\rho_2, m_\pi, r) \right]^2\nonumber \\
V_\sigma(r) &=& \sum_{i=1}^3 \alpha_i^{(\sigma)} e^{-\frac{r^2}{{\beta_i}^2}}, \ \ \
 V_\tau(r) = \sum_{i=1}^3 \alpha_i^{(\tau)} e^{-\frac{r^2}{{\beta_i}^2}} \nonumber \\
V_{\sigma \tau}(r) &=& \sum_{i=1}^3 \alpha_i^{(\sigma \tau)} e^{-\frac{r^2}{{\beta_i}^2}} + \lambda_1^{(\sigma \tau)} {\cal{Y}}(\rho_1, m_\pi, r).
\label{EQ:pot_rot}
\end{eqnarray}
The relation between the parameters are imposed as 
\begin{eqnarray}
\left( \begin{array}{c} 
\alpha_i({^{11}S_0}) \\ 
\alpha_i({^{31}S_0}) \\ 
\alpha_i({^{13}S_1}) \\ 
\alpha_i({^{33}S_1})  
\end{array} \right)
= 
\hat{A} 
\left( \begin{array}{c} 
\alpha_i^{(0)} \\ 
\alpha_i^{(\sigma)} \\ 
\alpha_i^{(\tau)} \\ 
\alpha_i^{(\sigma \tau)}  
\end{array} \right), \hspace*{3mm}
\left( \begin{array}{ccc} 
\lambda_1({^{11}S_0}) \\ 
\lambda_1({^{31}S_0}) \\ 
\lambda_1({^{13}S_1}) \\ 
\lambda_1({^{33}S_1})  
\end{array} \right)
= 
\left( \begin{array}{c} 
 9\lambda_1^{(\sigma \tau)} \\ 
-3\lambda_1^{(\sigma \tau)} \\ 
-3\lambda_1^{(\sigma \tau)} \\ 
  \lambda_1^{(\sigma \tau)}  
\end{array} \right), 
\label{EQ:prm_rot}
\end{eqnarray}
and  $\lambda_2(C) = \lambda_2^{(0)}$ being independent of the channel, $C$.

It is in order here to mention about the fitting procedure of the 
off-diagonal $\Lambda \Lambda$-$N \Xi$ potential.
  Although there is nothing wrong 
   to solve the coupled-channel Schr\"{o}dinger equation with non-Hermitian potential,  it is customary to use  Hermitian 
   potential in phenomenological studies in nuclear physics.   Since 
   the difference between $V^{\Lambda \Lambda}_{N \Xi}$ and $V^{N \Xi}_{\Lambda \Lambda}$   
    are confined only at short distances  if any (see  Fig.~\ref{FIG_NXS0I0} (b,c)), it does not
    affect the low-energy scattering observables.  We have checked this explicitly by choosing
   an Hermitian potential with the off-diagonal part is taken either $V^{\Lambda \Lambda}_{N \Xi}$,
     $V^{N \Xi}_{\Lambda \Lambda}$  or    their average $\bar{V}^{\Lambda \Lambda}_{N \Xi}
     =(V^{\Lambda \Lambda}_{N \Xi}+V^{N \Xi}_{\Lambda \Lambda})/2$. As shown in \ref{APP.HERM},
    the phase shifts  in these three cases do not have difference within the statistical errors.     
   Therefore, in the following we show the fit parameters corresponding to $\bar{V}^{\Lambda \Lambda}_{N \Xi} $.

Final fit parameters are given in Table~\ref{TAB.fit_params2} for 
 $V^{\Lambda \Lambda}({^{11}S_0})$ and Table~\ref{TAB.fit_params3} for
   $\bar{V}^{\Lambda \Lambda}_{N\Xi} ({^{11}S_0})$, with three different values $t/a= 11, 12, 13$.
  Also shown in Table~\ref{TAB.fit_params} are those for 
  $V^{N\Xi}$ in ${^{11}S_0}$, ${^{31}S_0}$, ${^{13}S_1}$ and  ${^{33}S_1}$ channels 
  with $t/a= 11, 12, 13$,
    where the data in all $N\Xi$ channels are fitted simultaneously.
    We perform uncorrelated fit for the potential,
      where the fit range is taken to be  $r=[0,2]$ fm independent of the potentials
    so that there are 220 coordinate data points to be fitted in each channel.
  For convenience, we show the corresponding parameters for $V^{N\Xi}$ in the operator basis
    in Table~\ref{TAB.fit_params_op}.
  As for the choice of these $t/a$, see \ref{APP.TDEP}.

\begin{table}[h]
\caption{Fitted parameters for $V^{\Lambda \Lambda}({^{11}S_0})$ with statistical errors using the data for $r=[0, 2]$fm.
 $\alpha_i$, $\beta_i$ and $\rho_i$ are given in units of [MeV], [fm] and [fm], respectively.
 $\lambda_1$ and $\lambda_2$ are given in units of [MeV $\cdot$ fm] and [MeV $\cdot$ fm${^2}$], respectively. 
The values of $\chi^2$/d.o.f. with \#d.o.f = $214$ are $1.30(40)$, $0.76(18)$ and $0.74(30)$ for $t/a=11$, $12$ and $13$, respectively.
  \label{TAB.fit_params2}}
\centering
\begin{tabular}{l|ll|ll|ll}
\hline \hline
 & \multicolumn{2}{l|}{Gauss-1} & \multicolumn{2}{l|}{Gauss-2} & \multicolumn{2}{l}{[Yukawa]${^2}$} \\
$t/a$ & $\alpha_1^{\Lambda \Lambda}$ & $\beta_1^{\Lambda \Lambda}$ & $\alpha_{2}^{\Lambda \Lambda}$ & $\beta_{2}^{\Lambda \Lambda}$ & $\lambda_2^{\Lambda \Lambda}$ & $\rho_2^{\Lambda \Lambda}$ \\
\hline 
$11$ & 1466.4(28.4) & 0.160(5) & 407.1(43.9) & 0.366(18) & -170.3(32.2) & 0.918(87) \\
$12$ & 1486.7(46.5) & 0.156(7) & 418.2(64.6) & 0.367(25) & -160.0(50.8) & 0.929(148) \\
$13$ & 1338.0(89.5) & 0.143(10) & 560.7(124.2) & 0.322(27) & -176.2(114.9) & 1.033(292) \\
\hline \hline
\end{tabular}
\end{table}
\begin{table}[ht]
\caption{Fitted parameters for the  transition potential  $\bar{V}^{\Lambda \Lambda}_{N\Xi} ({^{11}S_0})$
  with the statistical errors using the data for $r=[0, 2]$fm. Units are the same as those in  Table \ref{TAB.fit_params2}.
  The values of $\chi^2$/d.o.f. with \#d.o.f = $214$ are $1.24(35)$, $1.01(28)$ and $1.13(28)$ for $t/a=11$, $12$ and $13$, respectively.
\label{TAB.fit_params3}}
\centering
\begin{tabular}{l|ll|ll|ll}
\hline \hline
 & \multicolumn{2}{l|}{Gauss-1} & \multicolumn{2}{l|}{Gauss-2} & \multicolumn{2}{l}{Yukawa} \\
$t/a$ & $\alpha_1$ & $\beta_1$ & $\alpha_2$ & $\beta_2$ & $\lambda_1$ & $\rho_1$ \\
\hline 
$11$ & 1228.0(21.9) & 0.187(7) & 294.9(16.6) & 0.433(16) &	-69.7(16.2) &	0.130(8) \\
$12$ & 1206.7(27.0) & 0.191(12) & 307.4(30.3) & 0.438(25) & -75.2(26.9) & 0.133(13) \\
$13$ & 1252.7(47.6) & 0.187(25) & 306.9(46.4) & 0.428(58) & -65.8(58.1) & 0.128(28) \\
\hline \hline
\end{tabular}
\end{table}
\begin{table}[h]
\caption{Fitted parameters  for 
  $V^{N\Xi}$ $({^{11}S_0}, {^{31}S_0}, {^{13}S_1}, {^{33}S_1})$ 
 with the statistical errors using the data for $r=[0, 2]$fm.
In the last column, $\lambda_2(C)$ is independent of the channel, $C$.
 Units are the same as those in  Table \ref{TAB.fit_params2}.
 For the corresponding parameters in the operator basis, see Table~\ref{TAB.fit_params_op}.
 The values of $\chi^2$/d.o.f.
   with \#d.o.f = 861 
   are $2.34(39)$, $1.38(21)$ and $1.28(18)$ for $t/a=11$, $12$ and $13$, respectively.
 \label{TAB.fit_params}}
\centering
\begin{tabular}{l|l|l|l|l|l}
\hline \hline
 & Gauss-1 & Gauss-2 & Gauss-3 & Yukawa & [Yukawa]${^2}$ \\
$t/a=11$ & $\alpha_1$ & $\alpha_2$ & $\alpha_3$ & $\lambda_1$ & $\lambda_2$ \\
\hline 
${^{11}S_0}$ & 40.2(36.1) & 51.5(28.2) & 30.5(14.9) & -14.6(1.6) & -109.8(7.9) \\
${^{31}S_0}$ & 1766.1(75.6) & 920.3(56.8) & 240.5(31.1) & 4.9(5) & -109.8(7.9) \\
${^{13}S_1}$ & 493.3(30.9) & 300.8(22.9) &  92.0(17.3) & 4.9(5) & -109.8(7.9) \\
${^{33}S_1}$ & 944.8(46.8) & 568.6(29.8) & 190.3(25.0) & -1.6(2) & -109.8(7.9) \\
\hline
& $\beta_1$ & $\beta_2$ & $\beta_3$ & $\rho_1$ & $\rho_2$ \\
\cline{2-6}
      & 0.129(3) & 0.258(12) & 0.569(21) & 0.249(38) & 0.609(23) \\
\hline \hline
\end{tabular}
\\ \vspace*{2mm}
\begin{tabular}{l|l|l|l|l|l}
\hline \hline
 & Gauss-1 & Gauss-2 & Gauss-3 & Yukawa & [Yukawa]${^2}$ \\
$t/a=12$ & $\alpha_1$ & $\alpha_2$ & $\alpha_3$ & $\lambda_1$ & $\lambda_2$ \\
\hline 
${^{11}S_0}$ & -81.3(54.3) & 171.1(59.1) & 4.9(27.3) & -12.8(2.2) & -97.3(9.6) \\
${^{31}S_0}$ & 1677.2(90.1) & 991.3(62.7) & 290.8(43.2) & 4.3(7) & -97.3(9.6) \\
${^{13}S_1}$ & 449.2(52.5) & 348.9(31.8) & 110.3(22.3) & 4.3(7) & -97.3(9.6) \\
${^{33}S_1}$ & 849.5(53.4) & 653.9(32.7) & 210.8(35.9) & -1.4(2) & -97.3(9.6) \\
\hline 
& $\beta_1$ & $\beta_2$ & $\beta_3$ & $\rho_1$ & $\rho_2$ \\
\cline{2-6}
     & 0.124(3) & 0.241(12) & 0.533(22) & 0.136(22) & 0.603(48) \\
\hline \hline
\end{tabular}
\\ \vspace*{2mm}
\begin{tabular}{l|l|l|l|l|l}
\hline \hline
 & Gauss-1 & Gauss-2 & Gauss-3 & Yukawa & [Yukawa]${^2}$ \\
$t/a=13$ & $\alpha_1$ & $\alpha_2$ & $\alpha_3$ & $\lambda_1$ & $\lambda_2$ \\
\hline 
${^{11}S_0}$ & 62.4(125.4) & -43.6(144.7) & 123.8(110.2) & -12.5(2.4) & -83.5(14.6) \\
${^{31}S_0}$ & 1599.4(308.3) & 879.8(324.3) & 496.7(136.8) & 4.2(8) & -83.5(14.6) \\
${^{13}S_1}$ & 345.5(106.5) & 287.0(153.7) & 268.9(120.9) & 4.2(8) & -83.5(14.6) \\
${^{33}S_1}$ & 836.0(163.3) & 487.2(213.1) & 383.1(125.2) & -1.4(3) & -83.5(14.6) \\
\hline
& $\beta_1$ & $\beta_2$ & $\beta_3$ & $\rho_1$ & $\rho_2$ \\
\cline{2-6}
      & 0.124(10) & 0.228(34) & 0.499(33) & 0.307(307) & 0.417(74) \\
\hline \hline
\end{tabular}
\end{table}

The analytic forms of the potential, Eqs. (\ref{eq:YD}), (\ref{eq:YS}) and (\ref{V(C)}), together  with the parameters in Table 2-4 are 
useful for phenomenological applications such as the 
 calculation of the scattering phase shifts (as given in Fig.~\ref{FIG2} and Fig.~\ref{FIG4}) 
 and also the calculation of the binding energy of $\Xi$ hypernuclei (as discussed in ref.\cite{Hiyama:2019kpw}).
  The diagonal potentials (Fig.\ref{FIG_NXS0I0}(a), Fig.\ref{FIG_NXS0I0}(d) and Fig.\ref{FIG_NXpot_lat}) are composed of the short-range part  (parametrized by 
   $\alpha_i$ and $\beta_i$) and the  medium/long range part (parametrized by $\lambda_i$, $\rho_i$).
   The positive values of $\alpha_i$ 
   correspond to short-range repulsion, while the negative   values of $\lambda_i$  correspond to medium/long range
   attraction.  

\section{Scattering observables}
\label{sec:scatt}

The $\Lambda\Lambda$ and $N\Xi$ coupled-channel scattering phase shifts in the ${^{11}S_0}$ 
channel are calculated by solving the coupled-channel Schr\"odinger equation in the infinite volume with the fitted potentials 
given in the previous section.  Since we consider low-energy scatterings,   we adopt  the  non-relativistic kinematics hereafter.

\begin{figure}[t]
  \begin{tabular}{c}
    \begin{subfigure}{0.48\textwidth}
      \includegraphics[scale=0.35, bb=0 0 390 285]{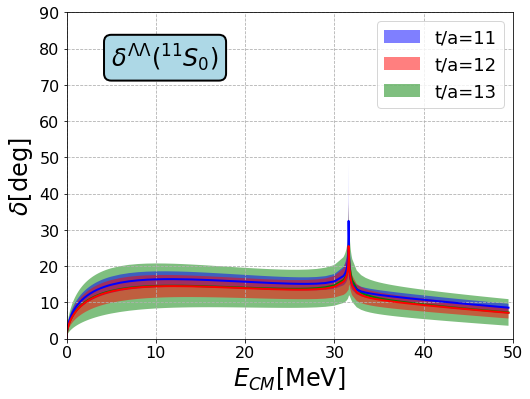}
      \caption{}
    \end{subfigure} 
    \begin{subfigure}{0.48\textwidth}
      \includegraphics[scale=0.35, bb=0 0 390 285]{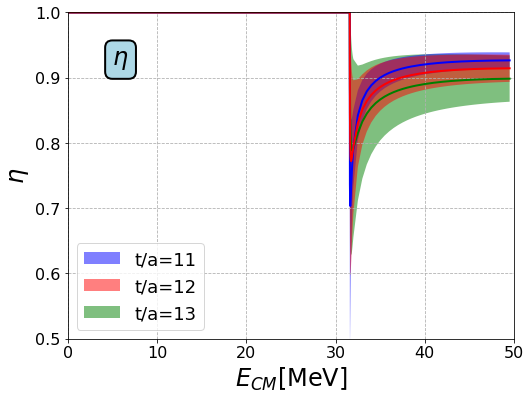}
      \caption{}
    \end{subfigure} \\
    \begin{minipage}{1\textwidth}
    \vspace*{14mm}
    \end{minipage} \\
    \begin{subfigure}{0.48\textwidth}
      \includegraphics[scale=0.35, bb=0 0 390 285]{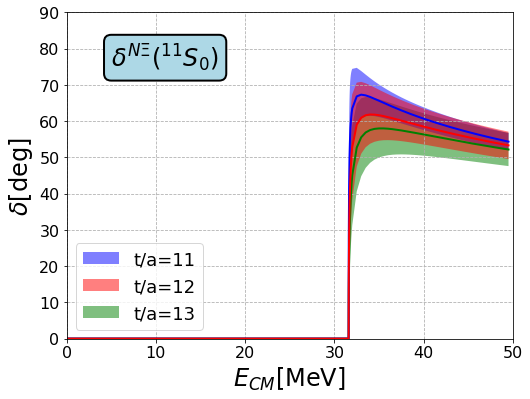}
      \caption{}
    \end{subfigure} 
  \end{tabular}
\caption{(a) $\Lambda \Lambda$ scattering phase shift, (b) $\Lambda \Lambda$ inelasticity, and 
 (c) $N\Xi$ scattering phase shift in the ${^{11}S_0}$ channel. \label{FIG2}}
\end{figure}

The $\Lambda \Lambda$ phase shifts and the inelasticity are 
 defined by the $\Lambda \Lambda$-component of the two-by-two S-matrix, $({\bf S})_{\Lambda\Lambda} = \eta \exp ( 2 i \delta_{\Lambda \Lambda})$.
  In Fig.~\ref{FIG2} (a,b), they are shown as a function of the center-of-mass energy  $E_{\rm CM}=k^2/m_{\Lambda} $ 
  with $k$ being the relative momentum between $\Lambda$s 
  for $t/a=11,12,13$.  The $t$-dependence is minor  within the statistical errors.
  We found that $\Lambda\Lambda$  attraction is rather weak, as inferred from Fig.~\ref{FIG_NXS0I0} (a).
  Accordingly,  no bound or resonant di-hyperon exits  around the $\Lambda \Lambda$ threshold in (2+1)-flavor QCD at nearly physical quark masses.
   This is  in contrast to the case of a possible $H$-dibaryon in 3-flavor QCD at heavy quark masses~\cite{Inoue:2011ai,Beane:2010hg}.
 
Low-energy part of  $\Lambda \Lambda$ phase shifts in Fig.~\ref{FIG2} (a) provides
 the scattering length and the effective range 
 using  the $S$-wave  effective range expansion (ERE) formula,
 \begin{eqnarray}
k \cot \delta = - \frac{1}{a_0} + \frac{1}{2}r_{\rm{eff}}k^2 + {\cal{O}}(k^4),
\end{eqnarray}
where we use the sign convention of $a_0$ in nuclear and atomic physics.
The results are
\begin{eqnarray}
a_0^{(\Lambda\Lambda)} = - 0.81 \pm 0.23{^{+0.00}_{-0.13}}\  [{\rm{fm}}], \hspace*{1em}
r^{(\Lambda\Lambda)}_{\rm{eff}} = 5.47 \pm 0.78{^{+0.09}_{-0.55}}\  [{\rm{fm}}] ,
\label{eq:ar}
\end{eqnarray}
where the central values and the statistical errors are estimated at
$t/a = 12$, while the systematic errors 
are estimated from the central values for $t/a = 11$ and $13$.
 For comparison, the experimental neutron-neutron ERE parameters are
$(a_0^{(nn)}, r^{(nn)}_{\rm{eff}} ) = ( -18.5, 2.80)$ fm.
 Our results in Eq.~(\ref{eq:ar})  were recently  confirmed to be consistent with a constraint obtained from 
 the  $\Lambda\Lambda$ momentum  correlation of  in p-p and p-Pb collisions \cite{Acharya:2018gyz}.
 (Note that the sign convention of  $a_0$  in \cite{Acharya:2018gyz}  is defined to be opposite from ours.) 

We note that the $\Lambda \Lambda$ phase shift in Fig.~\ref{FIG2} (a)  and the inelasticity $\eta$
 in Fig.~\ref{FIG2} (b) near the $N\Xi$ threshold, show a sharp enhancement and a rapid drop 
 and show an  enhancement, respectively, due to the off-diagonal coupling.
  Also,  Fig.~\ref{FIG2} (c) shows  a sharp increase of the  $N\Xi$  phase shift $\delta^{N\Xi}$ 
 up to about $60^\circ$ just above the $N\Xi$ threshold, which indicates a significant $N\Xi$ attraction
  in the ${^{11}S_0}$ channel.  Indeed,  we have confirmed 
   that the $N\Xi$ system is  in the unitary region and  a virtual pole is created   in the ${^{11}S_0}$ channel.

\begin{figure}[t]
  \begin{tabular}{c}
    \begin{subfigure}{0.48\textwidth}
      \includegraphics[scale=0.32, bb=0 0 390 285]{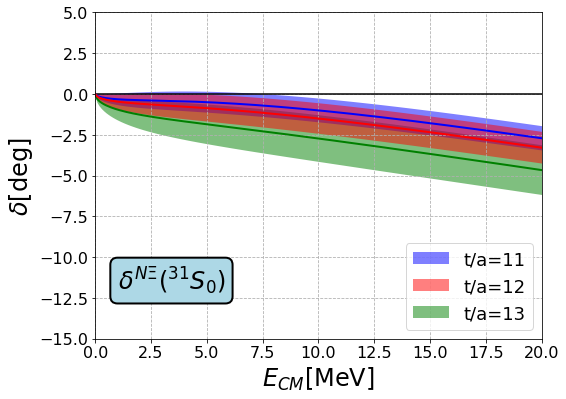}
      \caption{}
    \end{subfigure} 
    \begin{subfigure}{0.48\textwidth}
    \end{subfigure} \\
    \begin{minipage}{1\textwidth}
    \vspace*{14mm}
    \end{minipage} \\
    \begin{subfigure}{0.48\textwidth}
      \includegraphics[scale=0.32, bb=0 0 390 285]{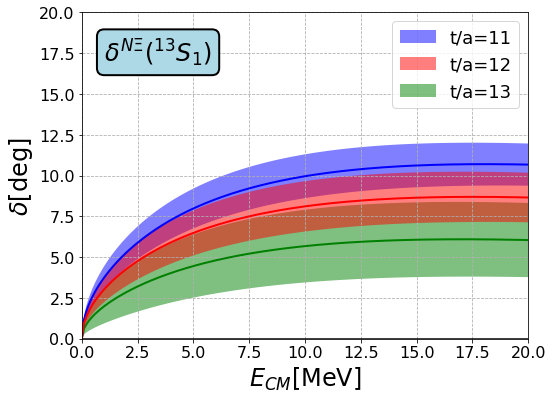}
      \caption{}
    \end{subfigure} 
    \begin{subfigure}{0.48\textwidth}
      \includegraphics[scale=0.32, bb=0 0 390 285]{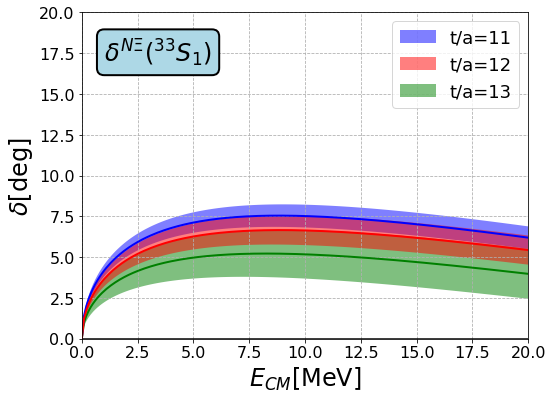}
      \caption{}
    \hspace*{1mm}
    \end{subfigure}
  \end{tabular}
\caption{The $N \Xi$ scattering phase shifts in ${^{31}S_0}$, ${^{13}S_1}$ and ${^{33}S_1}$ are shown in (a), (b) and (c), respectively.
 \label{FIG4}}
\end{figure}

The $S$-wave $N \Xi$ scattering phase shifts in ${^{31}S_0}$, ${^{13}S_1}$ and ${^{33}S_1}$ channel are shown in Fig.~\ref{FIG4} as a function of $E_{\rm CM}=k^2 \times  (1/(2m_N) + 1/ (2m_{\Xi}) )$. 
We find that the interaction in  the ${^{31}S_0}$ channel is weakly repulsive 
 while the ${^{13}S_1}$ and ${^{33}S_1}$ channels are weakly attractive, at low energies.

\section{Summary and conclusion remarks}
\label{sec:sum}
We have studied strangeness ${\cal{S}}=-2$ baryon-baryon interactions focusing on the $S$-wave $\Lambda\Lambda$ and
 $N \Xi$ potentials using  the (2+1)-flavor lattice QCD configurations at the almost physical point ($m_\pi \simeq 146{\rm{MeV}}$ and $m_K \simeq 525{\rm{MeV}}$)  analyzed by  the coupled-channel HAL QCD method.
  Resultant lattice QCD potentials in different isospin-spin channels (${^{11}S_0}$, ${^{31}S_0}$, ${^{13}S_1}$ and ${^{33}S_1}$)
   are parametrized by analytic functions (a combination of Gaussian and Yukawa forms)  for calculating 
    the scattering observables such as the phase shift and inelasticity.
    
We found that  $\Lambda \Lambda$ (${^{11}S_0}$) has attraction at low-energies, while it is not strong enough to generate bound or resonant dihyperon around the  $\Lambda \Lambda$ threshold.
  Our  scattering length and the effective range in  Eq.~(\ref{eq:ar}) were recently confirmed to be consistent with  an experimental constraint by ALICE experiment at LHC \cite{Acharya:2018gyz}.
 On the other hand,  we found that
  the $N \Xi ({^{11}S_0}$) has relatively a strong attraction to drive the system into the unitary regime,
   while  $N \Xi ({^{31}S_0}$) is  weakly repulsive and
  $N \Xi({^{13}S_1})$ and   $N \Xi({^{33}S_1})$ are weakly attractive.
  These features may lead to a light $\Xi$ hypernuclei as recently discussed in \cite{Hiyama:2019kpw}.
  Also, they introduce an attractive momentum correlation between proton and $\Xi^{-}$ on top of the 
   Coulomb attraction as suggested in \cite{Hatsuda:2017uxk} and confirmed recently by ALICE experiment
   at LHC \cite{Acharya:2019sms}. 
    
    There remain several future problems to be solved.  First of all, we need to carry out (2+1)-flavor
    and (1+1+1)-flavor lattice simulations exactly at the physical point to check whether the virtual pole
    in the $N\Xi ({^{11}S_0}$) channel turns into a resonance below the $N\Xi $ threshold.
    Another issue is to carry out full channel coupling analysis with $\Lambda \Sigma$ and 
     $\Sigma\Sigma$ to cover the scattering energy beyond the $\Lambda \Sigma$ threshold.

\section*{Acknowledgements}
We thank members of PACS Collaboration for the gauge configuration generations. 
The lattice QCD calculations have been performed on the K computer at RIKEN (hp120281, hp130023,hp140209, hp150223, hp150262, hp160211, hp170230), HOKUSAI FX100 computer at RIKEN (G15023, G16030, G17002) and HA-PACS at University of Tsukuba (14a-20, 15a-30). 
We thank ILDG/JLDG \cite{JLDGILDG1,JLDGILDG2,JLDGILDG3} which serves as an essential infrastructure in this study. 
We thank the authors of cuLGT code \cite{Schrock:2012fj} for the gauge fixing. 
This work is supported in part by the Grant-in-Aid of the Japanese Ministry of Education, Sciences and Technology, Sports and Culture (MEXT)
for Scientific Research (Nos. JP16H03978, JP18H05236, JP18H05407, JP19K03879), by SPIRE (Strategic Program for Innovative REsearch), by “Priority Issue on Post-K computer” (Elucidation of the Fundamental Laws and Evolution of the Universe) and by Joint Institute for Computational Fundamental Science (JICFuS). 

\newpage

\appendix
\setcounter{figure}{0}
\section{$t$-dependence of $N \Xi$ potentials} 
\label{APP.TDEP}

\begin{figure}
  \begin{tabular}{c}
    \begin{subfigure}{0.48\textwidth}
      \includegraphics[scale=0.34, bb=0 0 390 285]{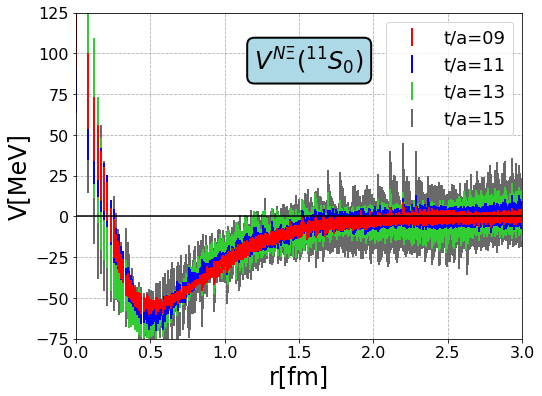}
      \caption{}
    \end{subfigure} 
    \begin{subfigure}{0.48\textwidth}
      \includegraphics[scale=0.34, bb=0 0 390 285]{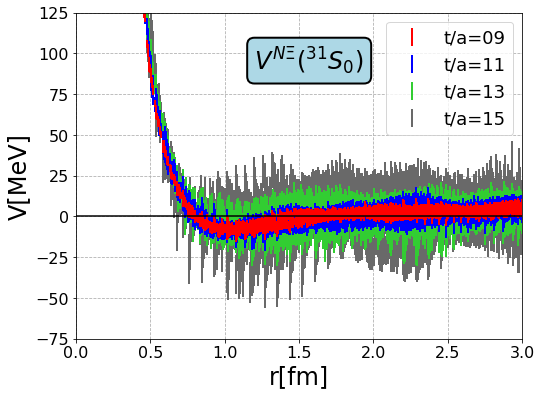}
      \caption{}
    \end{subfigure} \\
    \begin{minipage}{1\textwidth}
    \vspace*{14mm}
    \end{minipage} \\
    \begin{subfigure}{0.48\textwidth}
      \includegraphics[scale=0.34, bb=0 0 390 285]{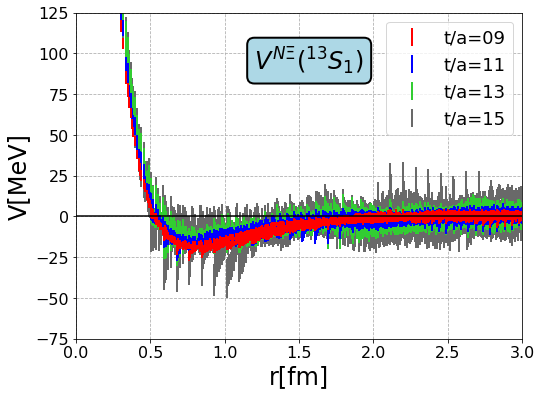}
      \caption{}
    \end{subfigure} 
    \begin{subfigure}{0.48\textwidth}
      \includegraphics[scale=0.34, bb=0 0 390 285]{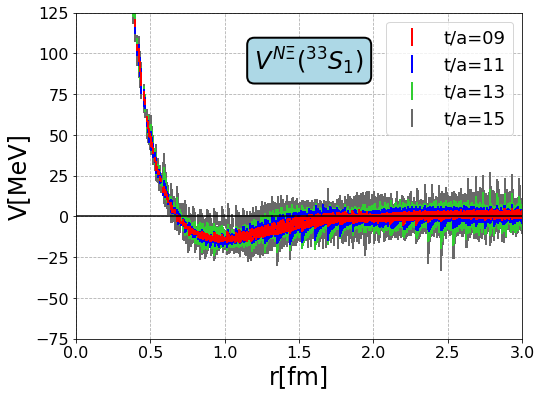}
      \caption{}
    \end{subfigure}
  \end{tabular}
\caption{The $t$-dependence of $S$-wave $N \Xi$ potentials. The $V^{N \Xi}({^{11}S_0})$, $V^{N \Xi}({^{31}S_0})$, $V^{N \Xi}({^{13}S_1})$ and $V^{N \Xi}({^{33}S_1})$ are shown in (a), (b), (c) and (d), respectively.  \label{FIG_TDEP}}
\end{figure}

 Fig.~\ref{FIG_TDEP} shows the  $t$-dependence of $S$-wave $N \Xi$ potentials in the range
  $t/a=9$ to $15$ which is wider than that used in the text.
   Overall stability of the results in this wider range of $t/a$ can be seen within the statistical error bars.
  Nevertheless, we find that the fitting by analytic functions in Sec.~\ref{sec:fit} is rather unstable against at short time
   (e.g. $t/a \le 10$) probably due to the inelastic state contaminations. Also the large statistical errors  prevent us to extract sensible fit parameters for $t/a \ge 14$.
  This is why we chose the optimal range $11 \le t/a \le 13$ throughout this paper.

\section{Dependence on different off-diagonal potentials}
\label{APP.HERM}

\setcounter{figure}{0}

\begin{figure}
      \includegraphics[scale=0.4, bb=0 0 693 199]{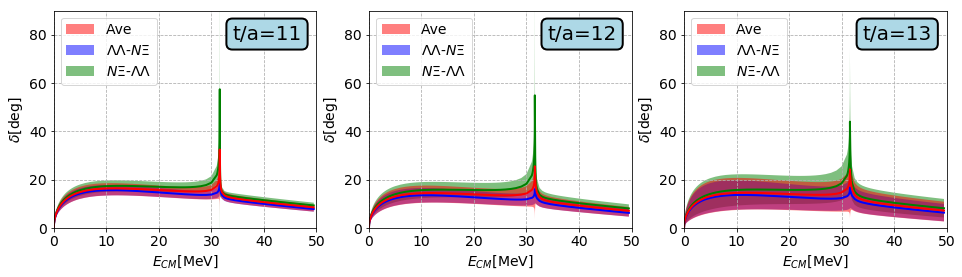}
\caption{The $\Lambda \Lambda$ scattering phase shifts for $t/a=11$ (left), $t/a=12$ (middle) and $t/a=13$ (right).
Blue, green and red curves correspond to the phase shifts calculated with $V^{\Lambda \Lambda}_{N \Xi}$, $V^{N \Xi}_{\Lambda \Lambda}$ and 
$\bar{V}^{\Lambda \Lambda}_{N \Xi}$, respectively.
 \label{FIG_LL_PHASE_HERM}}
\end{figure}

\begin{figure}
\vspace{1cm}
      \includegraphics[scale=0.4, bb=0 0 693 199]{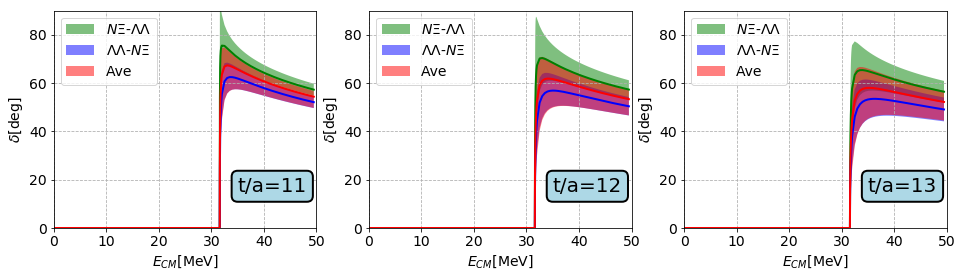}
\caption{Same with Fig.~\ref{FIG_LL_PHASE_HERM} but for $N \Xi$ scattering phase shifts.
 \label{FIG_NX_PHASE_HERM}}
\end{figure}

To check the observable difference among three choices of the off-diagonal part of the Hermitian potential
   ($V^{\Lambda \Lambda}_{N \Xi}$, $V^{N \Xi}_{\Lambda \Lambda}$ and their average $\bar{V}^{\Lambda \Lambda}_{N \Xi}$),
  the $\Lambda \Lambda$ scattering phase shifts and the $N \Xi$ scattering phase shifts
   are shown in  Fig.~\ref{FIG_LL_PHASE_HERM} and in  Fig.~\ref{FIG_NX_PHASE_HERM}, respectively.
   Within the statistical errors,  three results are consistent with each other for all $t/a$ used in this paper.
   Thus we consider $\bar{V}^{\Lambda \Lambda}_{N \Xi}$ in the text.

\newpage
\section{Fit parameters of $N \Xi$ potentials in the operator basis}
\label{APP.fit.op_basis}

\setcounter{table}{0}

We show the fit parameters of $N \Xi$ potentials in the operator basis,
$\left( V_{0}, V_{\sigma}, V_{\tau}, V_{\sigma \tau} \right)$ given in Eq.~(\ref{eq:op_basis}).
For parameters in the spin-isospin basis, see Table~\ref{TAB.fit_params}.

\begin{table}[h]
\caption{Fitting parameters  for 
  $V^{N\Xi}$
  in the operator basis
 with the statistical errors.
 Units are the same as those in  Table \ref{TAB.fit_params}.
 \label{TAB.fit_params_op}}
\centering
\begin{tabular}{l|l|l|l|l|l}
\hline \hline
 & Gauss-1 & Gauss-2 & Gauss-3 & Yukawa & [Yukawa]${^2}$ \\
$t/a=11$ & $\alpha_1$ & $\alpha_2$ & $\alpha_3$ & $\lambda_1$ & $\lambda_2$ \\
\hline 
$V_0$ & 957.6(44.7) & 552.0(29.3) & 171.3(23.6) & --- & -109.8(7.9) \\
$V_\sigma$ & -125.7(8.3) & -50.4(6.5) & -5.6(1.6) & --- & --- \\
$V_\tau$ & 192.5(9.9) & 104.5(8.0) & 31.6(3.1) & --- & --- \\
$V_{\sigma \tau}$ & -79.6(2.7) & -37.6(2.8) & -7.0(0.9) & -1.6(2) & --- \\
\hline
& $\beta_1$ & $\beta_2$ & $\beta_3$ & $\rho_1$ & $\rho_2$ \\
\cline{2-6}
      & 0.129(3) & 0.258(12) & 0.569(21) & 0.249(38) & 0.609(23) \\
\hline \hline
\end{tabular}
\\ \vspace*{2mm}
\begin{tabular}{l|l|l|l|l|l}
\hline \hline
 & Gauss-1 & Gauss-2 & Gauss-3 & Yukawa & [Yukawa]${^2}$ \\
$t/a=12$ & $\alpha_1$ & $\alpha_2$ & $\alpha_3$ & $\lambda_1$ & $\lambda_2$ \\
\hline 
$V_0$ & 871.4(54.6) & 629.8(33.0) & 194.1(33.0) & --- & -97.3(9.6) \\
$V_\sigma$ & -122.0(9.3) & -52.2(8.9) & -8.4(2.7) & --- & --- \\
$V_\tau$ & 185.0(10.0) & 108.5(7.2) & 36.7(5.2) & --- & --- \\
$V_{\sigma \tau}$ & -84.9(5.7) & -32.2(5.4) & -11.6(1.6) & -1.4(2) & --- \\
\hline 
& $\beta_1$ & $\beta_2$ & $\beta_3$ & $\rho_1$ & $\rho_2$ \\
\cline{2-6}
     & 0.124(3) & 0.241(12) & 0.533(22) & 0.136(22) & 0.603(48) \\
\hline \hline

\end{tabular}
\\ \vspace*{2mm}
\begin{tabular}{l|l|l|l|l|l}
\hline \hline
 & Gauss-1 & Gauss-2 & Gauss-3 & Yukawa & [Yukawa]${^2}$ \\
$t/a=13$ & $\alpha_1$ & $\alpha_2$ & $\alpha_3$ & $\lambda_1$ & $\lambda_2$ \\
\hline 
$V_0$ & 838.9(158.0) & 490.1(206.4) & 366.8(123.9) & --- & -83.5(14.6) \\
$V_\sigma$ & -125.4(27.1) & -53.0(24.1) & -12.2(4.4) & --- & --- \\
$V_\tau$ & 188.0(40.5) & 95.2(36.6) & 44.7(8.4) & --- & --- \\
$V_{\sigma \tau}$ & -65.4(20.6) & -45.2(17.2) & -16.2(4.7) & -1.4(3) & --- \\
\hline
& $\beta_1$ & $\beta_2$ & $\beta_3$ & $\rho_1$ & $\rho_2$ \\
\cline{2-6}
      & 0.124(10) & 0.228(34) & 0.499(33) & 0.307(307) & 0.417(74) \\
\hline \hline

\end{tabular}
\end{table}
%

\bibliography{basename of .bib file}

\end{document}